\documentclass[showpacs,preprintnumbers,amsmath,amssymb]{revtex4}
\usepackage{amsmath,amssymb,graphics,epsfig,subfigure}
\usepackage{color}

\begin{document}
\renewcommand{\baselinestretch}{1.3}

\title{Photon orbits and thermodynamic phase transition of $d$-dimensional charged AdS black holes}

\author{Shao-Wen Wei \footnote{weishw@lzu.edu.cn},
        Yu-Xiao Liu \footnote{liuyx@lzu.edu.cn}}

\affiliation{Institute of Theoretical Physics, Lanzhou University, Lanzhou 730000, People's Republic of China}

\begin{abstract}
We study the relationship between the null geodesics and thermodynamic phase transition for the charged AdS black hole. In the reduced parameter space, we find that there exist non-monotonic behaviors of the photon sphere radius and the minimum impact parameter for the pressure below its critical value. The study also shows that the changes of the photon sphere radius and the minimum impact parameter can serve as order parameters for the small-large black hole phase transition. In particular, these changes have an universal exponent of $\frac{1}{2}$ near the critical point for any dimension $d$ of spacetime. These results imply that there may exist universal critical behavior of gravity near the thermodynamic critical point of the black hole system.
\end{abstract}



\maketitle

\section{Introduction}

Recently, there is an extensive study on the black hole thermodynamics in the extended phase space with the cosmological constant being treated as pressure~\cite{Caldarelli1,Cvetic,Lu,Kastor,Dolan,Dolan2,Dolan3}. Interestingly, Kubiznak and Mann observed the thermodynamic stable small-large black hole phase transition for the four-dimensional charged AdS black holes~\cite{Kubiznak}, which mimics qualitatively the behavior of the Van der Waals (vdW) fluid. Such small-large black hole phase transition can be traced back to Refs.~\cite{Chamblin,Chamblin2,Caldarelli,Peca}. Shortly thereafter, the study was generalized to other black hole systems, and some abundant phase structures were discovered, such as the reentrant phase transition, triple point, isolated critical point, and superfluid black hole phase~\cite{Gunasekaran,Altamirano,Mann,Frassino,Wei0,Kostouki,Wei1,Hennigar,ZouYue,
Hendi,Hendi2,Hendi3,Hendi4,Momeni,Chakraborty} (for a recent review see~\cite{Teo} and references therein).

Probing the small-large black hole phase transition is an interesting subject. In Ref.~\cite{Liu}, the authors found that there is a dramatic change in the slope of the quasinormal modes (QNMs) before and after the black hole's thermodynamic phase transition. This provides us a dynamic probe for the phase transition. The study was also extended to other AdS black holes ~\cite{Mahapatra,Chabab,Zou2,Prasia,www}. These results imply that at low temperature or pressure, the phase transition can be effectively probed by measuring the change of the slope of the QNM frequencies. While at high temperature or pressure, extra measurement should be introduced, such as the non-monotonic behaviors of the imaginary part of the QNM frequencies~\cite{www}.

On the other hand, we would like to study whether there exist some relationships between the gravity and thermodynamics for the black hole system. In Refs.~\cite{Cvetic2,Cvetic3}, the authors pointed out that the presence of photon orbits signals a possible York-Hawking-Page type phase transition. Using the stable photon orbit, the effect of a negative cosmological constant on the instabilities of black hole spacetimes was investigated in Ref.~\cite{Tang}.

From the side of gravity, the null geodesics plays a center and important role in the black hole physics. In particularly, the photon sphere has a close relation with the strong gravitational effects near the black hole, such as the lensing, shadow, as well as the gravitational waves~\cite{Cardoso,Stefanov,WeiLiuGuo,HodHod,WeiLiu3,Raffaelli,Franzin,Stuchlik}. In general, there are two key quantities correspond to the photon sphere. One is the radius of the photon sphere, and another is the impact parameter related to the photon sphere radius. In this paper, we would like to examine the behavior of the photon sphere radius and the impact parameter when the black hole phase transition takes place. Certainly, it is also worth investigating the critical phenomena. These results will shed some new lights on the critical behavior of gravity near the critical point of the black hole thermodynamic phase transition.

This work is organized as follows. In Sec.~\ref{null}, we study the geodesics for the charged AdS black hole. And the photon sphere radius and the impact parameter are introduced. In Sec.~\ref{fouradsbh}, we obtain the photon sphere radius and the impact parameter for the four-dimensional black hole case. In the reduced parameter space, both of them are found to have the non-monotonic behavior when the phase transition occurs. Moreover, the changes of the radius and the impact parameter during the phase transition can serve as the order parameters to describe the black hole phase transition. The critical exponent of these changes near the critical point are also obtained. This result is generalized to higher-dimensional black hole case in Sec.~\ref{highadsbh}. Finally, the conclusions and discussions are presented in Sec.~\ref{Conclusion}.

\section{Null geodesics of charged AdS black hole}
\label{null}

The solution for a charged static and spherically symmetric black hole in
$d(\geq 4)$-dimensional spacetime is
\begin{eqnarray}
 ds^{2}&=&-f(r)dt^{2}+\frac{1}{f(r)}dr^{2}+r^{2}d\Omega_{(d-2)}^{2},\label{metric}\\
 A_t&=&-\sqrt{\frac{d-2}{2(d-3)}}\frac{q}{r^{d-3}},
\end{eqnarray}
where $d\Omega_{(d-2)}^{2}$ is the line element on the unit
$(d-2)$-dimensional sphere $S^{(d-2)}$ with the usual angular coordinates $\theta_{i}\in [0,\;\pi]$ $(i=1,\cdots,d-3)$ and $\phi\in [0,\;2\pi]$, and the metric function $f(r)$ reads
\begin{eqnarray}
 f(r)=1-\frac{m}{r^{d-3}}+\frac{q^{2}}{r^{2(d-3)}}+\frac{16\pi P r^{2}}{(d-1)(d-2)}.
\end{eqnarray}
Here the pressure $P$ is related to the cosmological constant $P=-\Lambda/8\pi$, and the parameters $m$ and $q$, respectively, correspond to the black hole mass $M$ and charge $Q$ as
\begin{eqnarray}
 M=\frac{d-2}{16\pi}\omega_{d-2}m,\quad
 Q=\frac{\sqrt{2(d-2)(d-3)}}{8\pi}\omega_{d-2}q,
\end{eqnarray}
where $\omega_{d-2}=\frac{2\pi^{(d-1)/2}}{\Gamma((d-1)/2)}$ is the volume of the unit sphere $S^{(d-2)}$. The black hole temperature and entropy read
\begin{eqnarray}
 T&=&\frac{1}{4\pi (d-2) r_{h}}\left(16\pi r_{h}^2
      \left(P-\frac{(d-2)(d-3)}{2}q^{2}r_{h}^{2(2-d)}\right)
     +(d-5)d+6\right),\label{tep}\\
 S&=&\frac{1}{4}~{\omega_{d-2}~r_{h}^{d-2}}.
\end{eqnarray}
It is known that this $d$-dimensional charged AdS black hole exhibits a vdW like phase transition, with the critical point given by~\cite{Gunasekaran}
\begin{eqnarray}
 T_{c}&=&\frac{4(d-3)^{2}}{(d-2)(2d-5)\pi v_{c}}\times q^{-\frac{1}{d-3}},\quad \nonumber  \\
 P_{c}&=&\frac{(d-3)^{2}}{(d-2)^{2}\pi v_{c}^{2}}\times q^{-\frac{2}{d-3}},\quad \label{critical} \\
 S_{c}&=&\frac{\omega_{d-2}}{4}\left(\frac{(d-2)}{4}v_{c}\right)^{d-2}\times q^{\frac{d-2}{d-3}}, \nonumber
\end{eqnarray}
where the critical specific volume is given by
\begin{eqnarray}
  v_{c}=\frac{4}{d-2}\left((d-2)(2d-5)\right)^{1/(2d-6)}.
\end{eqnarray}
It is clear that this critical point closely depends on the black hole charge.  Moreover, we would like to show the thermodynamic quantities in the reduced parameter space, where a reduced quantity is defined as $\tilde{A}=\frac{A}{A_{c}}$. Then the reduced temperature or state equation will of the form
\begin{eqnarray}
 \tilde{T}=\frac{\tilde{S}^{\frac{5-2d}{d-2}}}{4(d-2)(d-3)}\times
 \left[-1+(2d-5)\tilde{S}^{2}\left((d-3)\tilde{P}+(d-2)\tilde{S}^{-\frac{2}{d-2}}\right)\right].
\end{eqnarray}
Interestingly, there is no charge parameter in the reduced temperature.

\begin{figure}
\centerline{\includegraphics[width=8cm]{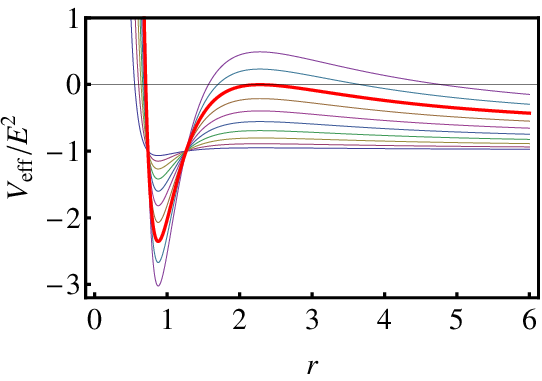}}
\caption{The effective potential for the four-dimensional charged AdS black hole. The parameters are set to $S$=5, $Q$=1, and $P$=0.003. The angular momentum $L/E$ of the photon varies from 0.8 to 4.4 from bottom to top (in the right side). The red thick line has the critical angular momentum $L_{c}/E\thickapprox$3.6.} \label{PVeff}
\end{figure}

Now, we would like to consider a free photon orbiting around a black hole on the equatorial hyperplane defined by $\theta_{i}=\frac{\pi}{2}$ for $i=1,\cdots ,d-3$.
Then the Lagrangian is
\begin{eqnarray}
 2\mathcal{L}=-f(r)\dot{t}^{2}+\dot{r}^{2}/f(r)+r^{2}\dot{\phi}^{2}.
 \label{lagrangian}
\end{eqnarray}
The dot over a symbol denotes the ordinary differentiation with respect to an affine parameter. From this Lagrangian, the generalized momentum can be calculated with $p_{\mu}=\frac{\partial \mathcal{L}}{\partial \dot{x}^{\mu}}=g_{\mu\nu}\dot{x}^{\nu}$, which gives
\begin{eqnarray}
 p_{t}   &=&-f(r)\dot{t}\equiv -E=\text{const},\label{pt}\\
 p_{\phi}&=&r^{2}\dot{\phi}\equiv L=\text{const},\label{phi}\\
 p_{r}   &=&\dot{r}/f(r),
\end{eqnarray}
where $E$ and $L$ denote the energy and orbital angular momentum of the photon, respectively. Solving Eqs.~(\ref{pt}) and (\ref{phi}), we easily get the $t$ motion and $\phi$ motion
\begin{eqnarray}
 \dot{t}&=&\frac{E}{f(r)},\\
 \dot{\phi}&=&\frac{L}{r^{2}\sin^{2}\theta}.\label{phit2}
\end{eqnarray}
The Hamiltonian for this system is
\begin{eqnarray}
 2\mathcal{H}&=&2(p_{\mu}\dot{x}^{\mu}-\mathcal{L})\nonumber\\
     &=&-f(r)\dot{t}^{2}+\dot{r}^{2}/f(r)+r^{2}\dot{\phi}^{2}\nonumber\\
     &=&-E\dot{t}+L\dot{\phi}+\dot{r}^{2}/f(r)=0.
\end{eqnarray}
With the help of the $t$ motion and $\phi$ motion, we can obtain the radial $r$ motion, which can be expressed as
\begin{eqnarray}
 \dot{r}^{2}+V_{eff}=0.\label{rt2}
\end{eqnarray}
The effective potential is
\begin{eqnarray}
 V_{eff}=\frac{L^{2}}{r^{2}}f(r)-E^{2}.\label{veff}
\end{eqnarray}
As an example, we plot the effective potential in Fig.~\ref{PVeff} for the four-dimensional charged AdS black hole with fixed parameters $S$=5, $Q$=1, and $P$=0.003. The orbital angular momentum $L$ of the photon varies from 0.8 to 4.4. Since $\dot{r}^{2}>$0, we require $V_{eff}<0$. So the photon can only appear in the range of negative $V_{eff}$. For small $L$, the photon can fall into the black hole from a place with large $r$. However, for large $L$, the peak of the potential will above zero, then the photon will be reflected before it falls into the black hole. Between the two cases, there exists a critical case described by the red thick line. Its peak just approaches zero at about $r$=2.28. At that point, the photon has zero radial velocity. So the photon will round the black hole at that radial distance. For a spherically symmetric static black hole, it corresponds to the photon sphere. In this paper, we mainly discuss the relation between this photon sphere and the thermodynamic phase transition of the black hole. The circular unstable photon sphere is determined by
\begin{eqnarray}
 V_{eff}=0,\quad \frac{\partial V_{eff}}{\partial r}=0,\quad \frac{\partial^{2} V_{eff}}{\partial r^{2}}<0.
\end{eqnarray}
Solving the second equation, one can find that the radius $r_{ps}$ satisfies
\begin{eqnarray}
 2f(r_{ps})-r_{ps}\partial_{r}f(r_{ps})=0.\label{rps1}
\end{eqnarray}
For a given metric function $f(r)$, we can obtain $r_{ps}$. Then solving the first equation, the minimum impact parameter or the critical angular momentum of the photon is
\begin{eqnarray}
 u_{ps}=\frac{L_{c}}{E}=\frac{r}{\sqrt{f(r)}}\bigg|_{r_{ps}}.\label{uspp}
\end{eqnarray}
Consider a photon starting from a place far away from the black hole and then passing by it, one will find that $u_{ps}$ has a close relation with the deflection angle of the photon. This is just the phenomenon of the black hole lensing. For a photon of large impact parameter $u$, its deflection angle is small. However when $u$ gradually tends to $u_{ps}$, the deflection angle will get larger and larger, and will be unbounded when $u_{ps}$ is arrived. So the minimum impact parameter $u_{ps}$ plays an important role in the black hole lensing~\cite{Bozza}.

Therefore, for a photon sphere, $r_{ps}$ and $u_{ps}$ are two key quantities. In the following sections, we will investigate the detailed behaviors of $r_{ps}$ and $u_{ps}$ near the small-large black hole phase transition, and discuss whether these quantities carry the phase transition information.

%%%%%%%%%%%%%%%%%%%%%%%%%%%%%%%%%%
%%%%%%%%%%%%%%%%%%%%%%%%%%%%%%%%%%

\section{$d$=4-dimensional charged AdS black hole}
\label{fouradsbh}

\begin{figure}
\center{\subfigure[]{\label{pTrps}
\includegraphics[width=6cm]{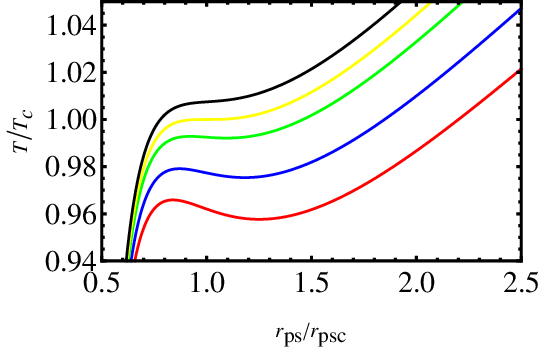}}
\subfigure[]{\label{pTups}
\includegraphics[width=6cm]{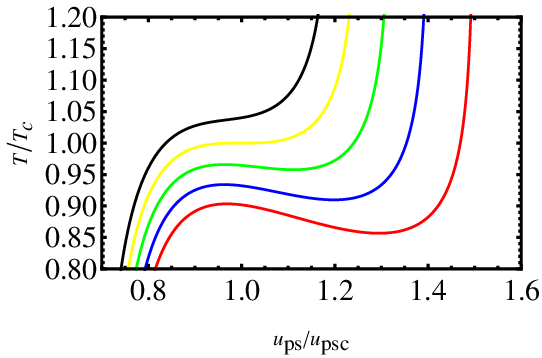}}}
\caption{(a) $\tilde{T}$ vs. $\tilde{r}_{ps}$ for fixed reduced pressure $\tilde{P}$=0.9, 0.94, 0.98, 1.0, and 1.02 from bottom to top. (b) $\tilde{T}$ vs. $\tilde{u}_{ps}$ for fixed $\tilde{P}$=0.7, 0.8, 0.9, 1.0, and 1.1 from bottom to top.}\label{ppTrps}
\end{figure}

For the four-dimensional charged AdS black hole, the temperature in the reduced parameter space is
\begin{eqnarray}
 \tilde{T}=\frac{3\tilde{P}\tilde{S}^{2}+6\tilde{S}-1}{8\tilde{S}^{3/2}}.
 \label{rted}
\end{eqnarray}
This state equation describes a vdW like phase transition. The coexistence curve of small and large black holes starts at $P$=0, and ends at a second-order critical point given in Eq.~(\ref{critical}). For the four-dimensional case, the critical point is~\cite{Kubiznak}
\begin{eqnarray}
 T_{c}=\frac{\sqrt{6}}{18\pi Q},\quad
 P_{c}=\frac{1}{96\pi Q^{2}},\quad
 S_{c}=6\pi Q^{2}.\label{cpc}
\end{eqnarray}
Fortunately, there exists an analytical form of the coexistence curve ~\cite{Spallucci,Lan}, which in the reduced parameter space reads
\begin{eqnarray}
 \tilde{T}^{2}=\frac{1}{2}{\tilde{P}\left(3-\sqrt{\tilde{P}}\right)}.
\end{eqnarray}
Let us turn to the null geodesic. For the four-dimensional case, we have the metric function
\begin{eqnarray}
 f(r)=1-\frac{2 M}{r}+\frac{Q^2}{r^2}+\frac{8}{3}\pi P r^2.
\end{eqnarray}
Solving Eq.~(\ref{rps1}), we obtain the radius of the photon sphere
\begin{eqnarray}
 r_{ps}=\frac{1}{2}(3M+\sqrt{9M^{2}-8Q^{2}}). \label{dd}
\end{eqnarray}
It is clear that this result is exactly equal to that of the asymptotically flat charged black hole, which may mean that the pressure $P$ does not affect the photon sphere. However, we need to note that the mass $M$ of the black hole depends on the pressure,
\begin{eqnarray}
 M=\frac{S(8PS+3)+3\pi Q^2}{6\sqrt{\pi S}}.
\end{eqnarray}
Plunging it into Eq.~(\ref{dd}), we will have the radius of the photon sphere
\begin{eqnarray}
 r_{ps}=\frac{3\pi Q^{2}+S(3+8PS)+\sqrt{(3\pi Q^{2}+S(3+8PS))^{2}-32\pi Q^{2}S}}{4\sqrt{\pi S}}.
\end{eqnarray}
It is clear that, for fixed charge $Q$ and entropy $S$, the radius $r_{ps}$ closely depends on the pressure $P$. At the critical point of the phase transition given in (\ref{cpc}), we have the critical radius
\begin{eqnarray}
 r_{psc}=(2+\sqrt{6})Q\approx4.45Q.
\end{eqnarray}
Then in the reduced parameter space, we get
\begin{eqnarray}
 \tilde{r}_{ps}\equiv\frac{r_{ps}}{r_{psc}}=\frac{3+3\tilde{S}(6+\tilde{P}\tilde{S})+\sqrt{3\left(3\tilde{P}^{2}\tilde{S}^{4}
    +36\tilde{P}\tilde{S}^{3}+6(18+\tilde{P})\tilde{S}^{2}-28\tilde{S}+3\right)}}
  {8(3+\sqrt{6})\sqrt{\tilde{S}}}.
\end{eqnarray}
Combining with (\ref{rted}), we can plot the reduced temperature $\tilde{T}$ as a function of the photon sphere radius $\tilde{r}_{ps}$ for fixed pressure $\tilde{P}$, which is shown in Fig.~\ref{pTrps}. From it, we can see that for the case $\tilde{P}<$1, there will be a non-monotonic behavior. One local maximum and one minimum are presented. While for the case $\tilde{P}>$1, the non-monotonic behavior disappears, and the reduced temperature $\tilde{T}$ is only a monotone increasing function of $\tilde{r}_{ps}$. For $\tilde{P}=1$, a deflection point can be found at $\tilde{r}_{ps}=1$. Such behavior of the reduced temperature is very similar to the isobar of the vdW fluid in the $\tilde{T}$-$\tilde{S}$ chart, which indicates that a first-order phase transition exists. By constructing the equal area law at the isobar, we can determine the phase transition point for the vdW fluid or for the black hole. For our case, one may apply the same technique in this $\tilde{T}$-$\tilde{r}_{ps}$ chart. A detailed analysis implies that the coexistence curve calculated by it approximately meets Eq.~(\ref{rted}). However, strictly speaking, one must obtain the coexistence curve from the free energy or by constructing appropriate equal area law.

%%%%%%%%%%%%%%%%%%%%%%%%%%%%%%%%%%%%%%%%%%%%%%%%%%%%%%%%%%%%%%%%%%%%%
%%%%%%%%%%%%%%%%%%%%%%%%%%%%%%%%%%%%%%%%%%%%%%%%%%%%%%%%%%%%%%%%%%%%%%%%%

On the other hand, plunging $r_{ps}$ into (\ref{uspp}), one can obtain the minimum impact parameter $u_{ps}$. However, it is in a complicated form and we will not show it. Nevertheless, we have the critical minimum impact parameter
\begin{eqnarray}
 u_{psc}=\frac{(3+\sqrt{6})^{2}Q}{\sqrt{19+8\sqrt{6}}}\approx4.78Q.
\end{eqnarray}
We also show the behavior of $\tilde{T}$ as a function of $\tilde{u}_{ps}$ in Fig.~\ref{pTups}. Interestingly, it also confirms a non-monotonic behavior for $\tilde{P}<1$. And that behavior disappears for $\tilde{P}\geq1$. Although such non-monotonic behavior in the $\tilde{T}$-$\tilde{r}_{ps}$ chart and $\tilde{T}$-$\tilde{u}_{ps}$ chart cannot be used to determined the thermodynamic small-large black hole phase transition by constructing the equal area laws, it does demonstrate that such non-monotonic behavior indicates the existence of the phase transition.

%%%%%%%%%%%%%%%%%%%%%%%%%%%%%%%%%%
\begin{figure}
\centerline{\includegraphics[width=6cm]{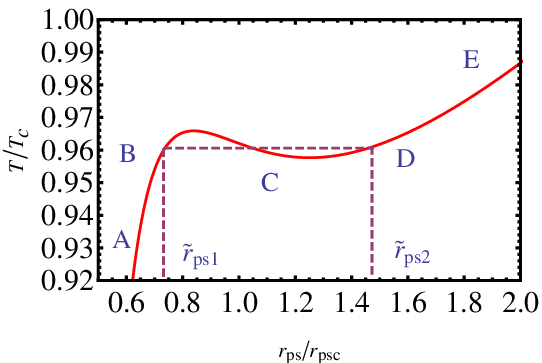}}
\caption{$\tilde{T}$ vs. $\tilde{r}_{ps}$ for fixed $\tilde{P}$=0.9. The horizontal line BD corresponds to the phase transition temperature $\tilde{T}$=0.9608.} \label{PTttrps}
\end{figure}
%%%%%%%%%%%%%%%%%%%%%%%%%%%%%%%%%%

\begin{figure}
\center{\subfigure[]{\label{pDeltar}
\includegraphics[width=6cm]{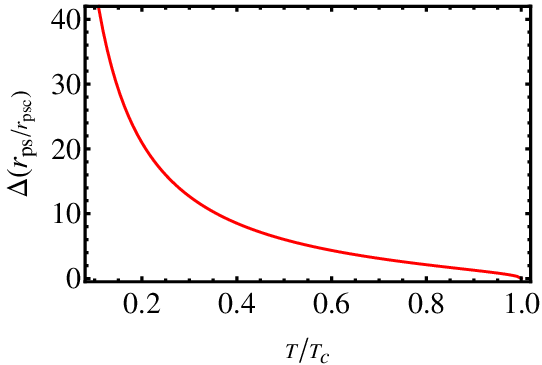}}
\subfigure[]{\label{pDeltau}
\includegraphics[width=6cm]{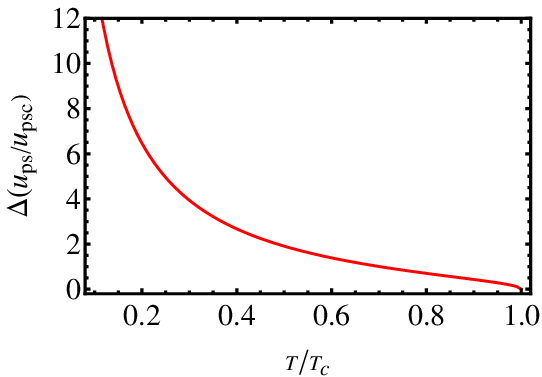}}}
\caption{Behaviors of $\Delta\tilde{r}_{ps}$ and $\Delta\tilde{u}_{ps}$ as a function of the phase transition temperature $\tilde{T}$. (a) $\Delta\tilde{r}_{ps}$ vs. $\tilde{T}$. (b) $\Delta\tilde{u}_{ps}$ vs. $\tilde{T}$.}\label{ppDeltar}
\end{figure}

In order to examine the changes of the photon sphere radius and the minimum impact parameter before and after the small-large black hole phase transition, we plot the reduced temperature as a function of the photon sphere radius with fixed $\tilde{P}$=0.9 in Fig.~\ref{PTttrps} for example. The horizontal line BD corresponds to the black hole phase transition with temperature $\tilde{T}$=0.9608, which is determined by Eq.~(\ref{rted}). With the increase of $\tilde{r}_{ps}$, the black hole phase transition occurs at the horizontal line BD. The photon sphere radius before and after the phase transition is $\tilde{r}_{ps1}=0.7385$ and $\tilde{r}_{ps2}=1.4669$. So, the photon sphere radius has a sudden change $\Delta\tilde{r}_{ps}$=0.7284 among the black hole phase transition. Increasing the temperature or pressure such that the critical point is approached, one will find $\tilde{r}_{ps1}=\tilde{r}_{ps2}=1$, which leads to $\Delta\tilde{r}_{ps}$=0. Thus if one observes a sudden change, then the black hole system must experience a first-order phase transition. The minimum impact parameter also possesses such behavior. We present the change of $\Delta\tilde{r}_{ps}$ and $\Delta\tilde{u}_{ps}$ in Fig.~\ref{ppDeltar}. From it, we can see that both $\Delta\tilde{r}_{ps}$ and $\Delta\tilde{u}_{ps}$ decrease with the phase transition temperature $\tilde{T}$, and approach to zero at $\tilde{T}$=1, where the first-order phase transition becomes a second-order one. Moreover, such pattern reminds us that $\Delta\tilde{r}_{ps}$ and $\Delta\tilde{u}_{ps}$ can serve as the order parameters for the black hole phase transition. It is also interesting to see whether they have the critical exponent near the critical point. For this purpose, we expend them near $\tilde{T}$=1, which gives
\begin{eqnarray}
 \Delta\tilde{r}_{ps}&\sim& 3.4641\times(1-\tilde{T})^{\frac{1}{2}}+\mathcal{O}((1-\tilde{T})^{\frac{3}{2}}),\\
 \Delta\tilde{u}_{ps}&\sim& 1.1981
 \times(1-\tilde{T})^{\frac{1}{2}}+\mathcal{O}((1-\tilde{T})^{\frac{3}{2}}).
\end{eqnarray}
This result exactly confirms that both $\Delta\tilde{r}_{ps}$ and $\Delta\tilde{u}_{ps}$ have the same critical exponent at the critical point. More importantly, the critical exponents of them are equal to $\frac{1}{2}$, which is the same as that of these order parameters from the thermodynamic side, such as the specific volume or number density~\cite{Weisw}. Thus this novel property strongly implies that there does exist a relationship between the photon orbits and thermodynamic phase transition of the black hole.

%%%%%%%%%%%%%%%%%%%%%%%%%%%%%%%%%%%%%%%%%%%%%%%%%%%%%%%%%%%%%%%%%%%%%
%%%%%%%%%%%%%%%%%%%%%%%%%%%%%%%%%%%%%%%%%%%%%%%%%%%%%%%%%%%%%%%%%%%%%%%%%

\section{$d\geq5$-dimensional charged AdS black hole}
\label{highadsbh}

For the higher-dimensional charged AdS black hole, there also exists the small-large black hole phase transition. Similar to the four-dimensional case, we can obtain the photon sphere radius $r_{ps}$ and the minimum impact parameter $u_{ps}$ for $d\geq5$ by solving Eqs.~(\ref{rps1}) and (\ref{uspp}). Since they have lengthy expressions, we will not show them. The critical values of $r_{ps}$ and $u_{ps}$ are listed in Table~\ref{tab1}. Both of them are proportional to $Q^{1/(d-3)}$. And these coefficients decrease with the dimension $d$ of spacetime.

%%%%%%%%%%%%%%%%%%%%%%%%%%%%%%%%%%%%%%%%%%%%%%%%%%%%%%%%%%%%%%%%%%%%%%%%%%%%
\begin{table}[h]
\begin{center}
\begin{tabular}{ccccccc}
  \hline\hline
  $d$ & 5 &6 &7 &8 &9 &10 \\\hline
  $r_{psc}/Q^{1/(d-3)}$ & 1.9701 & 1.5593 & 1.4135 & 1.3497 & 1.3213 & 1.3109\\ \hline
  $u_{psc}/Q^{1/(d-3)}$ & 1.6625 & 1.2065 & 1.0491 & 0.9788 & 0.9447 & 0.9285\\ \hline\hline
\end{tabular}
\caption{Critical values of $r_{ps}$ and $u_{ps}$.}\label{tab1}
\end{center}
\end{table}
%%%%%%%%%%%%%%%%%%%%%%%%%%%%%%%%%%%%%%%%%%%%%%%%%%%%%%%%%%%%%%%%%%%%%%%%%%%%%%

\begin{figure}
\center{\subfigure[~$\tilde{T}$ vs. $\tilde{r}_{ps}$]{\label{pTrps610}
\includegraphics[width=6cm]{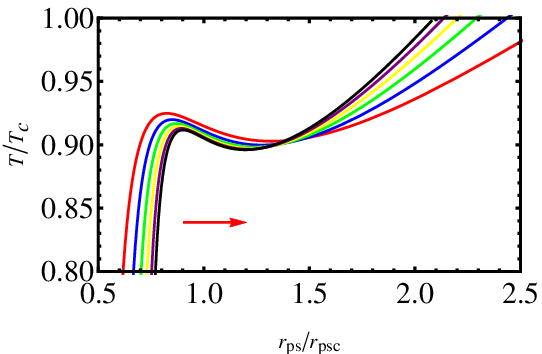}}
\subfigure[~$\tilde{T}$ vs. $\tilde{u}_{ps}$]{\label{pTups610}
\includegraphics[width=6cm]{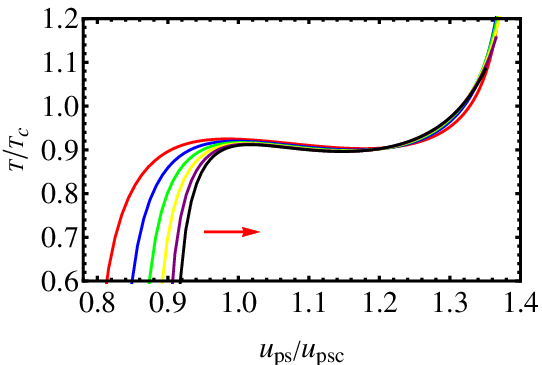}}}
\caption{(a) $\tilde{T}$ vs. $\tilde{r}_{ps}$ and (b) $\tilde{T}$ vs. $\tilde{u}_{ps}$ for fixed reduced pressure $\tilde{P}$=0.8, and the dimension of spacetime $d$=6-10. The arrows indicate the increase of $d$.}\label{ppTups610}
\end{figure}

\begin{figure}
\center{\subfigure[~$\Delta\tilde{r}_{ps}$ vs. $\tilde{T}$]{\label{pDeltar610}
\includegraphics[width=6cm]{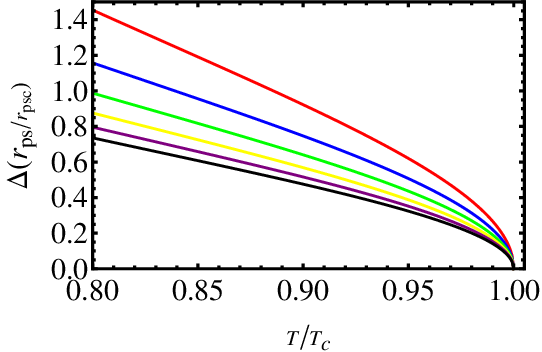}}
\subfigure[~$\Delta\tilde{u}_{ps}$ vs. $\tilde{T}$]{\label{pDeltau610}
\includegraphics[width=6cm]{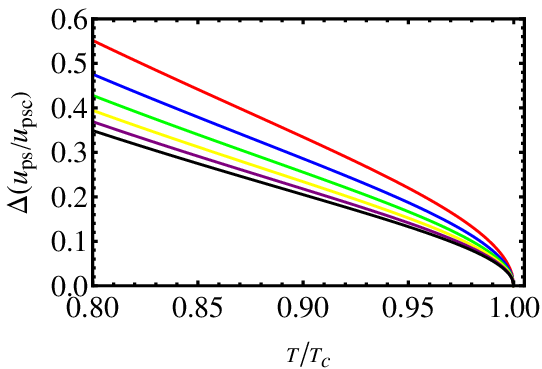}}}
\caption{Behaviors of $\Delta\tilde{r}_{ps}$ and $\Delta\tilde{u}_{ps}$ as functions of the phase transition temperature $\tilde{T}$. (a) $\Delta\tilde{r}_{ps}$ vs. $\tilde{T}$. (b) $\Delta\tilde{u}_{ps}$ vs. $\tilde{T}$. The dimension of spacetime is set to $d$=5-10 from top to bottom.}\label{ppDeltau610}
\end{figure}

In Fig.~\ref{ppTups610}, we plot the behavior of the reduced temperature $\tilde{T}$ as a function of $\tilde{r}_{ps}$ or $\tilde{u}_{ps}$ with fixed $\tilde{P}$=0.8 for $d$=5-10, respectively. These curves obviously show the non-monotonic behavior, which indicates a first-order phase transition. In fact, the small-large black hole phase transition does exist for this case $\tilde{P}$=0.8. However, different from the $d$=4 case, there are no analytic forms of the coexistence curves for $d\geq5$. Fortunately, we have obtained the fitting formula of the coexistence curves by fitting the numerical data in Ref.~\cite{Wei1}, which is in the form
\begin{eqnarray}
 \tilde{P}=\sum^{10}_{i=0} a_{i}\tilde{T}^{i},\quad\tilde{T}\in(0,1).
\end{eqnarray}
It is worth to mention here that our fitting formula of the coexistence curve agrees with the numerical values to $10^{-7}$. By making use of this fitting formula, we can get the sudden changes of the $\tilde{r}_{ps}$ and $\tilde{u}_{ps}$. For clear, we show them in Fig.~\ref{ppDeltau610} as a function of the phase transition temperature. Similar to the four-dimensional case, $\Delta\tilde{r}_{ps}$ and $\Delta\tilde{u}_{ps}$ decrease with the phase transition temperature $\tilde{T}$, and approach to zero at $\tilde{T}$=1. So even in the higher-dimensional black hole case, $\Delta\tilde{r}_{ps}$ and $\Delta\tilde{u}_{ps}$ can serve as order parameters to describe the black hole phase transition. Moreover, we can also find that, for fixed temperature, both $\Delta\tilde{r}_{ps}$ and $\Delta\tilde{u}_{ps}$ decrease with the dimension of spacetime.

%%%%%%%%%%%%%%%%%%%%%%%%%%%%%%%%%%%%%%%%%%%%%%%%%%%%%%%%%%%%%%%%%%%%%
%%%%%%%%%%%%%%%%%%%%%%%%%%%%%%%%%%%%%%%%%%%%%%%%%%%%%%%%%%%%%%%%%%%%%%%%%

%%%%%%%%%%%%%%%%%%%%%%%%%%%%%%%%%%%%%%%%%%%%%%%%%%%%%%%%%%%%%%%%%%%%%
%%%%%%%%%%%%%%%%%%%%%%%%%%%%%%%%%%%%%%%%%%%%%%%%%%%%%%%%%%%%%%%%%%%%%%%%%

%%%%%%%%%%%%%%%%%%%%%%%%%%%%%%%%%%%%%%%%%%%%%%%%%%%%%%%%%%%%%%%%%%%%%%%%%%%%
\begin{table}[h]
\begin{center}
\begin{tabular}{cccccccc}
  \hline\hline
   &$d$ & 5 &6 &7 &8 &9 &10 \\\hline
$\Delta\tilde{r}_{ps}$
  &$a$ & 2.7249 & 2.2345 & 1.9207 & 1.7052 & 1.5464 & 1.4237\\
  &$\delta$ & 0.5048 & 0.5037 & 0.5028 & 0.5024 & 0.5023 & 0.5023\\ \hline
$\Delta\tilde{u}_{ps}$
  &$a$& 0.9560 & 0.8117 & 0.7209 & 0.6585 & 0.6112 & 0.5740\\
  &$\delta$ & 0.5067 & 0.5070 & 0.5069 & 0.5072 & 0.5073 & 0.5075\\ \hline\hline
\end{tabular}
\caption{Fitting coefficients of $a$ and $\delta$ for $\Delta\tilde{r}_{ps}$ and $\Delta\tilde{u}_{ps}$ near the critical temperature $\tilde{T}$=1.}\label{tab2}
\end{center}
\end{table}
%%%%%%%%%%%%%%%%%%%%%%%%%%%%%%%%%%%%%%%%%%%%%%%%%%%%%%%%%%%%%%%%%%%%%%%%%%%%%%

Also, we are expected to investigate the critical behavior of $\Delta\tilde{r}_{ps}$ and $\Delta\tilde{u}_{ps}$ near the critical temperature $\tilde{T}$=1. Here we fit the numerical coefficients of $\Delta\tilde{r}_{ps}$ and $\Delta\tilde{u}_{ps}$ to the following form
\begin{eqnarray}
 \Delta\tilde{r}_{ps},\;\Delta\tilde{u}_{ps}
  \sim a\times(1-\tilde{T})^{\delta}.
\end{eqnarray}
The fitting results are given in Table~\ref{tab2}, from which we can see that the values of the parameter $a$ for both $\Delta\tilde{r}_{ps}$ and $\Delta\tilde{u}_{ps}$ decrease with $d$. More importantly, we obtain the result that the values of the exponent parameter $b$ are all around $\frac{1}{2}$ with error smaller than 1.5\%. So the exponent $\delta$ is an universal parameter, which is consistent with the analytic value of the $d$=4 case. This result further confirms that there exists a relationship between the photon orbits and thermodynamic phase transition of the charged AdS black holes.

\section{Conclusions and discussions}
\label{Conclusion}

In this paper, we studied the relationship between the photon orbits and thermodynamic phase transition of the charged AdS black hole. The photon sphere radius $r_{ps}$ and minimum impact parameter $u_{ps}$ of the photon orbits were examined in detail.

By solving the null geodesics, the expressions of $r_{ps}$ and $u_{ps}$ were obtained. In the reduced parameter space, both $\tilde{r}_{ps}$ and $\tilde{u}_{ps}$ are charge independent. The small-large black hole phase transition also shares this interesting property. For some certain pressures below the critical case, we found that there appear the non-monotonic behaviors both in $\tilde{T}$-$\tilde{r}_{ps}$ and $\tilde{T}$-$\tilde{u}_{ps}$ charts. When the pressure is larger than its critical value, these non-monotonic behaviors disappear. So this suggests that there exists an obvious relationship between the thermodynamic phase transition and the non-monotonic behaviors of $\tilde{r}_{ps}$ and $\tilde{u}_{ps}$. This non-monotonic behavior also implies that there are sudden changes of $\tilde{r}_{ps}$ and $\tilde{u}_{ps}$ when the first-order black hole phase transition takes place. For any value of $d$, we found that the changes $\Delta\tilde{r}_{ps}$ and $\Delta\tilde{u}_{ps}$ decrease with the phase transition temperature, and vanish at the critical temperature. So they can serve as order parameters for the thermodynamic phase transition of the black hole.

Moreover, we also investigated the critical behavior of $\Delta\tilde{r}_{ps}$ and $\Delta\tilde{u}_{ps}$ near the critical temperature. For $d$=4, we obtained the analytic critical exponent $\delta=\frac{1}{2}$ for $\Delta\tilde{r}_{ps}$ and $\Delta\tilde{u}_{ps}$, which is the same as that of other order parameters from the thermodynamic side. On the other hand, for the higher-dimensional black hole, there is no the analytic value. However, with the fitting method, we found the values of the critical exponent $\delta$ for $d$=5-10 are approximately equal to $\frac{1}{2}$. So for any $d$-dimensional charged AdS black hole, $\Delta\tilde{r}_{ps}$ and $\Delta\tilde{u}_{ps}$ have a universal critical exponent. This further confirms the existence of the relationship between the photon orbits and thermodynamic phase transition of the black hole system.

Before ending this paper, we emphasize that thermodynamic phase transition is encoded in and can be revealed by the sudden change of the photon sphere radius and minimum impact parameter of the black hole. And we conjecture that this is a universal property for the black hole system. Next, it is worthwhile to generalize our study to other black hole systems. Moreover, since $r_{ps}$ and $u_{ps}$ are closely related to the gravity effects of the black hole, finding the relationship between the thermodynamic phase transition and some astronomical observables is also worthwhile to pursue.

\section*{Acknowledgements}
We would like to thank G. W. Gibbons for the useful correspondence.
This work was supported by the National Natural Science Foundation of China (Grants No. 11675064, No. 11522541, No. 11375075 and No. 11205074)).

\end{document}